\documentclass[aps,prd,showpacs,preprintnumbers]{revtex4}
\usepackage{amsmath,epsfig}
\usepackage{subfigure}
\usepackage{amssymb,amsfonts}
\usepackage{latexsym}
\usepackage{epsfig}
\usepackage{amssymb}
 \usepackage{bm}
\newcommand{\be}{\begin{eqnarray}}
\newcommand{\ee}{\end{eqnarray}}

\newcommand{\bea}{\begin{eqnarray}}
\newcommand{\eea}{\end{eqnarray}}

\begin{document}

\title{Neutral and charged (pseduo)scalar mesons and diquarks under magnetic fields}
\author{Hao Liu$^{1,2,3}$}
\author{Xinyang Wang$^{1}$}
\author{Lang Yu$^{4}$}
\author{Mei Huang$^{1,2,5}$}
\affiliation{$^{1}$  Institute of High Energy Physics, Chinese Academy of Sciences, Beijing 100049, P.R. China}
\affiliation{$^{2}$ School of Physics Sciences, University of Chinese Academy of Sciences, Beijing 100039, China}
\affiliation{$^{3}$ Jinyuan Senior High School, Shanghai 200333, China}
\affiliation{$^{4}$ Center of Theoretical Physics and College of Physics,Jilin University,Changchun 130012, People¡¯s Republic of China}
\affiliation{$^{5}$ Theoretical Physics Center for Science Facilities, Chinese Academy of Sciences, Beijing 100049, P.R. China}

\begin{abstract}
We investigate both (pseudo)scalar mesons and diquarks in the presence of external magnetic field in the framework of the two-flavored Nambu--Jona-Lasinio (NJL) model, where mesons and diquarks are constructed by infinite sum of quark-loop chains by using random phase approximation. The polarization function of the quark-loop is calculated to the leading order of $1/N_c$ expansion by taking the quark propagator in the Landau level representation. We systematically investigate the masses behaviors of scalar $\sigma$ meson, neutral and charged pions as well as the scalar diquarks, with respect to the magnetic field strength at finite temperature and chemical  potential. It is shown that the numerical results of both neutral and charged pions are consistent with the lattice QCD simulations. The mass of the charge neutral pion keeps almost a constant under the magnetic field, which is preserved by the remnant symmetry of QCD$\times$QED in the vacuum. The mass of the charge neutral scalar $\sigma$ is around two times quark mass and increases with the magnetic field due to the magnetic catalysis effect, which is an typical example showing that the polarized internal quark structure cannot be neglected when we consider the meson properties under magnetic field. For the charged particles,  the one quark-antiquark loop contribution to the charged $\pi^{\pm}$ increases essentially with the increase of magnetic fields due to the magnetic catalysis of the polarized quarks.  However, the one quark-quark loop contribution to the scalar diquark mass is negative comparing with the point-particle result and the loop effect is small.
\end{abstract}
\maketitle

\section{Introduction}
The influence of an external magnetic field on QCD vacuum and matter has attracted great attention in the past few decades (see Refs.~\cite{Andersen:2014xxa,Miransky:2015ava,Huang:2015oca}), since there are at least three high-energy physical systems where strong magnetic fields may play an important role. First, it's predicted by some cosmological models that extremely strong magnetic fields as high as $10^{20-23}$ G might be produced during the electroweak phase transition in the early universe~\cite{Vachaspati:1991nm}. Second, the magnetic fields on the surface of magnetars could reach $10^{14}$-$10^{15}$ G, while in the inner core of magnetars the magnitude of magnetic fields is expected to be on the order of $10^{18}$-$10^{20}$ G~\cite{Duncan:1992hi}. Finally, in the non-central heavy ion collisions, very strong but short-lived magnetic fields can be generated, of which the strength can reach up to $B\sim10^{18}$ G at Relativistic Heavy Ion Collider (RHIC) and $B\sim10^{20}$ G at the Large Hadron Collider (LHC) ~\cite{Skokov:2009qp,Deng:2012pc}. More importantly, heavy ion collisions provides a controllable experimental platform to investigate plenty of fascinating effects of strong magnetic fields on strongly interacting matter, for example, the chiral magnetic effect (CME)~\cite{Kharzeev:2007tn,Kharzeev:2007jp,Fukushima:2008xe}, the magnetic catalysis~\cite{Klevansky:1989vi,Klimenko:1990rh,Gusynin:1995nb} and inverse magnetic catalysis~\cite{Bali:20111213} effect, and the vacuum superconductivity~\cite{Chernodub:2010qx,Chernodub:2011mc}.

Hadron properties at finite magnetic field has also attracted much interests. For a free point-like charged particle in a static uniform external magnetic field $B$,  its energy level has the form of $\varepsilon_{n,s_z}^2(p_z) = p_z^2+(2 n - 2 \, \text{sign}(q) s_z + 1) |qB| + m^2$ with $q$ the electric charge of the particle, $n$ characterizing the Landau levels, $s_z$ the projection of particle's spin on the magnetic field axis $z$, and $p_z$ the particle's momentum along the magnetic field. For a point-like charged vector meson $\rho^\pm$, its mass $M_{\rho^\pm}(B) = \sqrt{m_{\rho^\pm}^2 - |eB|}$ decreases linearly with the magnetic filed to zero at  the critical magnetic field $eB_c = m_{\rho^\pm}^2\approx 0.6$ GeV$^2$ \cite{Chernodub:2010qx}, which indicates the instability of the ground state towards the condensation of the charged $\rho$ mesons in the vacuum. It is then checked in the NJL model, by considering the quark-loop corrections, the charged $\rho$ mass decreases to zero at a rather small critical magnetic field $eB_c \approx 0.2 {\rm GeV}^2 $ \cite{Li:2013aa}, which is only 1/3 of the results from the point-particle results. The magnetic field strength dependence of the $\rho^\pm$ meson mass has been widely investigated by different approaches~\cite{Chernodub:2010qx,Chernodub:2011mc,Callebaut:2011uc,Ammon:2011je,Hidaka:2012mz,Frasca:2013kka,Andreichikov:2013zba,Wang:phd,Liu:2014uwa,
Liu:2015pna,Liu:2016vuw,Kawaguchi:2015gpt,Luschevskaya:2014mna,Luschevskaya:2015bea,Zhang:2016qrl}, however, the existence of charged $\rho$ meson condensation in strong magnetic field is still under debate nowadays (see Refs~\cite{Hidaka:2012mz,Chernodub:2012zx,Li:2013aa,Chernodub:2013uja}). One of our motivations in this work is to extend our method for charged vector mesons in  \cite{Liu:2014uwa,Liu:2015pna,Liu:2016vuw} to less debated neutral and charged (pseudo)scalars, e.g., $\sigma$ and $\pi$ \cite{Bali:2015vua,Luschevskaya:2015cko,Hidaka:2012mz,
Andreichikov:2013zba,Fayazbakhsh:2012vr,Fayazbakhsh:2013cha,Avancini:2015ady,Simonov:2015xta,Avancini:2016fgq}, which has less debating result at finite magnetic field, and check whether our method in  \cite{Liu:2014uwa,Liu:2015pna,Liu:2016vuw} is valid.

On the other hand, the chiral symmetry breaking and restoration under a strong magnetic field is another significant issue of QCD, which is deeply related to hadrons' properties, such as the pion mass and the pion decay constant via the Gell¨CMann-Oakes-Renner (GOR) relation~\cite{GellMann:1968rz}. It means that, exploring the modification of hadrons' properties in the magnetized hot and/or dense medium, will help to understand the effects of magnetic fields on the chiral phase transition at finite temperature and chemical potential. Thus, the behaviors of the mass spectrum and weak decay constant of pions have then been extensively studied recently~\cite{Hidaka:2012mz,Andreichikov:2013zba,Fayazbakhsh:2012vr,Fayazbakhsh:2013cha,Avancini:2015ady,Simonov:2015xta,Avancini:2016fgq}. However, most of them only focused on the neutral pion due to the difficulty of treating different charged quark propagators under magnetic fields. Therefore, in this paper, we will study not only the pseudoscalar neutral pion and scalar $\sigma$ meson but also the charged pions in the two-flavored NJL model under magnetic fields. In addition, since the mass generation of nucleon is also an important feature of dynamical chiral symmetry breaking, we further analyze the scalar color $\bar{3}$ diquark channel in the magnetic field, which will help to probe the properties of nucleons in the magnetized medium in the future.

This paper is organized as following: in the next section, we give a general expression of the two-flavor NJL model including the scalar and pseudo-scalar channels, and then derive the polarization functions of both (pseudo)scalar meson channels and the scalar color $\bar{3}$ diquark channel in a magnetic field at finite temperature and chemical potential. In Sec.~\ref{Numerical} we show our numerical results and analysis. Finally, the discussion and conclusion part is given in Sec.~\ref{Conclusion}.

\section{FOMALISM}
\label{formalism}
\subsection{The two flavor NJL model in an external magnetic field}
We investigate the (pseudo)scalar mesons and the diquark in the scalar channel by using a low energy approximation of QCD effective model SU(2) NJL model under an external magnetic field, the Lagrangian is given by\cite{Nambu:1961tp,Nambu:1961fr}
\begin{eqnarray}
{\cal{L}}&=&\bar{\psi}(i\not{\!\!D}-\hat{m}+\mu\gamma^0)\psi+G_{S}\left[(\bar{\psi}\psi)^2
   +(\bar{\psi}i\gamma^5\vec{\tau}\psi)^2\right] -\frac{1}{4}F_{\mu\nu}F^{\mu\nu}.
\label{eq:L:basic}
\eea
Where $\psi$ represents the quark field of two light flavors $u$ and $d$, the current mass matrix $\hat{m}=\text{diag}(m_u,m_d)$, we can assume that $m_u=m_d=m_0$. $\tau^a=(I,\vec{\tau})$ with $\vec{\tau}=(\tau^1,\tau^2,\tau^3)$ corresponding  to the isospin Pauli matrix. $G_S$ is the coupling constants corresponding to the (pseudo)scalar channel. The covariant derivative, $D_{\mu}=\partial_{\mu}-i Q e A_{\mu}^{ext}$, couples quarks to an external magnetic field $\bm{B}=(0,0,B)$ along the positive $z$ direction via a background field, for example, $A_{\mu}^{ext}=(0,0,Bx,0)$ and the field strength tensor is defined by $F_{\mu\nu}=\partial_{[\mu}A_{\nu]}^{ext}$, $Q=(-1/3,2/3)$ is  a diagonal matrix in the flavor space which respects to the  electric charge of the quark field ($u,d$).

 Semi-bosonizing the above Lagrangian  and the  Eq.(\ref{eq:L:basic}) can be rewritten as
  \begin{eqnarray}\label{NE2b}
{\cal{L}}_{sb}&=&\bar{\psi}(x)\left(i\gamma^{\mu}D_{\mu}-\hat{m}+\mu\gamma^0\right)\psi(x)
-\bar{\psi}\left(\sigma+i\gamma_5\vec{\tau}\cdot\vec{\pi}\right)\psi-\frac{(\sigma^2+\vec{\pi}^2)}{4G_S}-\frac{B^2}{2},
\end{eqnarray}
where the Euler-Lagrange equation of motion for the auxiliary fields leads to the constraints as follows:
\begin{eqnarray}\label{NE3b}
\sigma(x)&=&-2G_S\left<\bar{\psi}(x)\psi(x)\right>,\\
\vec{\pi}(x)&=&-2G_S\left<\bar{\psi}(x)i\gamma_5\vec{\tau}\psi(x)\right>.
\end{eqnarray}
For each flavor, by introducing the chirality projector $P_L=\frac{1+\gamma^5}{2}$ and $P_R=\frac{1-\gamma^5}{2}$, we have
\begin{equation}
u =\left(\begin{array}{c}
                    u_L \\
                    u_R
                   \end{array}
            \right), \ \
\bar{u} =( \bar{u}_R \ \  \bar{u}_L ),
\end{equation}
\begin{equation}
d =\left(\begin{array}{c}
                    d_L \\
                    d_R
                   \end{array}
            \right), \ \
\bar{d} =( \bar{d}_R \ \  \bar{d}_L ),
\end{equation}
and for two-flavor spinor $\bar{\psi}=(\bar{u} \ \  \bar{d})$,  the auxiliary neutral scalar and pseudo scalar $\sigma$ and $\pi^0$ can be written as follows:
\begin{eqnarray}\label{neutralmeson-chirality}
\sigma& \sim & \bar{\psi}\psi= \bar{u}_R u_L +  \bar{d}_R d_L +  \bar{u}_L u_R + \bar{d}_L d_R, \\
\pi^0&\sim & \bar{\psi}i\gamma_5\tau^3\psi=i (\bar{u}_L u_R -  \bar{u}_R u_L) - i (\bar{d}_L d_R - \bar{d}_R d_L).
\end{eqnarray}
It is obviously to find that the neutral scalar $\sigma$ is symmetric in the flavor space, while the neutral $\pi^0$ is anti-symmetric in the flavor space and keeps as a pseudo Goldstone mode.
The charged $\pi^{\pm}$ can be represented as:
\begin{eqnarray}\label{chargedmeson-chirality}
\pi^+& \sim & \bar{\psi}i\gamma_5\tau^-\psi =\sqrt{2} i(\bar{d}_L u_R -  \bar{d}_R u_L ), \\
\pi^-&\sim & \bar{\psi}i\gamma_5\tau^+\psi= \sqrt{2} i (\bar{u}_L d_R -  \bar{u}_R d_L),
\end{eqnarray}
with $\tau^{\pm}=\frac{1}{\sqrt{2}}(\tau_1 \pm i \tau_2)$.

The quark-antiquark condensation which gives quark the dynamical mass and the constituent quark mass is obtained as
\be \label{eq:gapequationmq}
M=m_0-2G_S \left<\bar{\psi}\psi\right>,
\ee
We should minimize the effective potential in order to obtain the dynamical quark mass, i.e., the $\sigma$ condensation. The one-loop effective potential in this model is  given as follows:
\begin{eqnarray} \label{eq:effectivepotential}
 \Omega&=&\frac{\sigma^2}{4G_S}+\frac{B^2}{2}-3\sum_{q_f\in\{\frac{2}{3},-\frac{1}{3}\}}\frac{|q_feB|}{\beta}\sum_{p=0}^{+\infty}\alpha_p\int_{-\infty}^{+\infty}\frac{dp_3}{4\pi^2} \left\{\beta E_q+\ln \left(1+e^{-\beta(E_q+\mu)}\right)+\ln \left(1+e^{-\beta(E_q-\mu)}\right)\right\},\nonumber\\
 \end{eqnarray}
 where $\beta=\frac{1}{T}$,  $\alpha_p=2-\delta_{p,0}$ represents the spin degeneracy.

\subsection{The scalar meson $\sigma$}
In the framework of the NJL model, the meson is $\bar{q}q$ bound state or resonance, it can be obtained from the quark-antiquark scattering amplitude\cite{
He:1997gn,Rehberg:1995nr}.   The meson is  constructed by summing up infinite quark-loop chains in the random phase approximation(RPA), the quark loop of
the $\sigma$ meson polarization function is calculated to the leading order of $1/N_c$ expansion,  the one-loop polarization function of $\sigma$ meson $\Pi_{\sigma}(q_{\bot},q_{||})$ in the magnetic field takes the form of \cite{Klevansky:1992qe}
\bea\label{sigmapolar}
\Pi_{\sigma}(q_{\bot},q_{||})=-i\int\frac{d^4k}{(2\pi)^4}\text{Tr}[\widetilde{S}(k)\widetilde{S}(p)],
\eea
where $q_{\bot}=(0,q_1,q_2,0), q_{||}=(q_0,0,0,q_3)$. $p=k+q_{\sigma}$ corresponds to the momentum conservation and the Landau level representation of the quark propagator $\widetilde{S}(k)$ is given by\cite{Gusynin:1995nb,Chodos:1990vv}
\bea\label{quarkpropa}
\widetilde{S}_Q(k)&=& i\exp\left(-\frac{\mathbf{k}_{\bot}^2}{|QeB|}\right)\sum_{n=0}^{\infty}(-1)^n\frac{D_n(QeB,k)}{k_0^2-k_3^2-M^2-2|QeB|n},
\eea
with
\bea\label{Dn}
D_n(QeB,k)&=&(k^0\gamma^0-k^3\gamma^3+M)\Big[(1-i\gamma^1\gamma^2\mathrm{sign}(QeB))L_n\left(2\frac{\mathbf{k}_{\bot}^2}{|QeB|}\right)\nonumber\\
&&-(1+i\gamma^1\gamma^2\mathrm{sign}(QeB))L_{n-1}\left(2\frac{\mathbf{k}^2_{\bot}}{|QeB|}\right)\Big]+4(k^1\gamma^1+k^2\gamma^2)L_{n-1}^1\left(2\frac{\mathbf{k}^2_{\bot}}{|QeB|}\right).\nonumber\\
\eea
Here, $L_n^{\alpha}$ are the generalized Laguerre polynomials and $L_n=L_n^0$.  

In the rest frame of $\sigma$ meson, i.e.,~$q^{\mu}_{\sigma}=(M_{\sigma},\mathbf{0})$, by using the quark propagator in the Eq.(\ref{quarkpropa}), the polarization function of $\sigma$ is given as follows,
\bea\label{sigmapropa1}
&&\Pi_{\sigma}(q_{\bot},q_{||})=3i\int\frac{d^4k}{(2\pi)^4}\sum_{q_f=\frac{2}{3},-\frac{1}{3}}\text{exp}\left(-\frac{2k^2_{\bot}}{|q_f eB|}\right)\sum_{p,k=0}^{\infty}(-1)^{p+k}\nonumber\\
&&\frac{1}{(p_0^2-k_3^2-M^2-2|q_f eB|p)(k_0^2-k_3^2-M^2-2|q_f eB|k)}\nonumber\\
&&\times\Bigg{\{}8(p_{||}\cdot k_{||}+M^2)\bigg[L_k\left(2\frac{k^2_{\bot}}{|q_f eB|}\right)L_p\left(2\frac{k^2_{\bot}}{|q_f eB|}\right)+L_{k-1}\left(2\frac{k^2_{\bot}}{|q_f eB|}\right)L_{p-1}\left(2\frac{k^2_{\bot}}{|q_f eB|}\right)\bigg]\nonumber\\
&&-64k^2_{\bot}L^1_{k-1}\left(2\frac{k^2_{\bot}}{|q_f eB|}\right)L^1_{p-1}\left(2\frac{k^2_{\bot}}{|q_f eB|}\right)\Bigg{\}}\nonumber\\
&=&
6i\int\frac{dk_0dk_3}{(2\pi)^3}\sum_{q_f=\frac{2}{3},-\frac{1}{3}}\sum_{k=0}^{\infty}\frac{p_{||}\cdot k_{||}-2|q_f eB|k+M^2}{(p_0^2-k_3^2-M^2-2|q_f eB|p)(k_0^2-k_3^2-M^2-2|q_f eB|k)}|q_f eB|\alpha_k.\nonumber\\
\eea
Here, we have used the relations
\bea\label{deltalandau}
&&\int_{0}^{\infty}k_{\bot}dk_{\bot}\text{exp}\left(-\frac{2k_{\bot}^2}{|q_f eB|}\right)L_p\left(2\frac{k^2_{\bot}}{|q_f eB|}\right)L_k\left(2\frac{k^2_{\bot}}{|q_f eB|}\right)=\frac{|q_f eB|}{4}\delta_{p,k},\nonumber\\
&&\int_{0}^{\infty}k_{\bot}^3dk_{\bot}\text{exp}\left(-\frac{2k^2_{\bot}}{|q_f eB|}\right)L_{k-1}^1\left(2\frac{k^2_{\bot}}{|q_f eB|}\right)L_{p-1}^1\left(2\frac{k^2_{\bot}}{|q_f eB|}\right)=\frac{|q_f eB|^2k}{8}\delta_{p-1,k-1}.
\eea
 Moreover,
 \bea
 \Pi_{\sigma}(q_{\bot},q_{||})=6i\int\frac{dk_3}{(2\pi)^2}\sum_{q_f=\frac{2}{3},-\frac{1}{3}}\sum_{k=0}^{\infty}\left[I_1+(2M^2-\frac{M^2_{\sigma}}{2})I_2\right]|q_f eB|\alpha_k,
 \eea
 where the functions $I_1$ and $I_2$ are given in appendix \ref{Integralk0}.

We can get the mass of $\sigma$ meson by solving the gap equation~\cite{Klevansky:1992qe}
\bea\label{Gapsigma}
1-2G_S \Pi_{\sigma}(q_{\bot},q_{\parallel})=0.
\eea

\subsection{The pseudoscalar meson $\pi$ }

 For neutral $\pi^0$ meson, the one-loop polarization function is
 \bea
 \Pi_{\pi^0}(q_{\bot},q_{||})=-i\int\frac{d^4k}{(2\pi)^4}Tr[i\gamma_5\tau^3\widetilde{S}(k)i\gamma_5\tau^3\widetilde{S}(p)].
 \eea
Similarly, $p=k+q_{\pi^0}$ and in the rest frame of the $\pi^0$ meson, i.e., $q^{\mu}_{\pi^0}=(M_{\pi^0},\mathbf{0})$, the $\Pi_{\pi^0}$ is written as
\bea\label{polarneutrpi}
&&\Pi_{\pi^0}(q_{\bot},q_{||})=-3i\int\frac{d^4k}{(2\pi)^4}\sum_{q_f=-\frac{1}{3},\frac{2}{3}}\text{exp}\left(-\frac{2k^2_{\bot}}{|q_f eB|}\right)\sum_{p,k=0}^{\infty}(-1)^{p+k}\nonumber\\
&&\frac{1}{(p_0^2-k_3^2-M^2-2|q_f eB|p)(k_0^2-k_3^2-M^2-2|q_f eB|k)}\nonumber\\
&&\times\Bigg{\{}8(-p_{||}\cdot k_{||}+M^2)\left[L_k\left(2\frac{k^2_{\bot}}{|q_f eB|}\right)L_p\left(2\frac{k^2_{\bot}}{|q_f eB|}\right)+L_{k-1}\left(2\frac{k^2_{\bot}}{|q_f eB|}\right)L_{p-1}\left(2\frac{k^2_{\bot}}{|q_f eB|}\right)\right]\nonumber\\
&&+64k^2_{\bot}L^1_{k-1}\left(2\frac{k^2_{\bot}}{|q_f eB|}\right)L^1_{p-1}\left(2\frac{k^2_{\bot}}{|q_f eB|}\right)\Bigg{\}}\nonumber\\
&&=6i\int\frac{dk_0dk_3}{(2\pi)^3}\sum_{q_f=-\frac{1}{3},\frac{2}{3}}|q_f eB|\sum_{k=0}^{\infty}\frac{\alpha_k\delta_{p,k}(p_{||}\cdot k_{||}-2|q_f eB|k-M^2)}{(p_0^2-k_3^2-M^2-2|q_f eB|p)(k_0^2-k_3^2-M^2-2|q_f eB|k)} \nonumber\\
&&=6i\int\frac{dk_3}{(2\pi)^2}\sum_{q_f=\frac{2}{3},-\frac{1}{3}}\sum_{k=0}^{\infty}\left(I_1-\frac{M^2_{\pi^0}}{2}I_2\right)|q_f eB|\alpha_k.
\eea

For charged $\pi^{\pm}$ meson, the one-loop polarization function is (the notation is only for $\pi^{+}$ meson, and there is a similar notation for $\pi^{-}$ meson)
\bea
\Pi_{\pi^{+}}(q_{\bot},q_{||})&=&-i\int\frac{d^4k}{(2\pi)^4}\text{Tr}[i\gamma_5\tau^-\widetilde{S}(k)i\gamma_5\tau^+\widetilde{S}(p)],
\eea
where $p=k+q_{\pi^+}$. In the rest frame of $\pi^+$ meson, the polarization function is given by
\bea
&&\Pi_{\pi^+}(q_{\bot},q_{||})=-6i\int\frac{d^4k}{(2\pi)^4}\text{exp}\left(-\frac{9k_{\bot}^2}{2|eB|}\right)\sum_{k=0,p=0}^{\infty}(-1)^{p+k}\nonumber\\
&&\frac{1}{(p_0^2-k_3^2-\frac{4}{3}|eB|p-M^2)(k_0^2-k_3^2-\frac{2}{3}|eB|k-M^2)}\nonumber\\
&&\times\Bigg{\{}8(p_{||}\cdot k_{||}-M^2)\bigg[L_p\left(2\frac{k_{\bot}^2}{|q_u eB|}\right)L_{k-1}\left(2\frac{k_{\bot}^2}{|q_d eB|}\right)+L_{p-1}\left(2\frac{k_{\bot}^2}{|q_u eB|}\right)L_{k}\left(2\frac{k_{\bot}^2}{|q_d eB|}\right)\bigg]\nonumber\\
&&+64k_{\bot}^2L_{p-1}^1\left(2\frac{k_{\bot}^2}{|q_u eB|}\right)L_{k-1}^1\left(2\frac{k_{\bot}^2}{|q_d eB|}\right)\Bigg{\}}\nonumber\\
&&=-6i\int\frac{d^3\mathbf{k}}{(2\pi)^3}\text{exp}(-\frac{9k_{\bot}^2}{2|eB|})\sum_{k=0,p=0}^{\infty}(-1)^{p+k}\Bigg{\{}\bigg[4(I_1'+I_1'')+4(2|q_d eB|k+2|q_u eB|p\nonumber\\
&&-M_{\pi^+}^2)I_2'\bigg]\bigg[L_p\left(2\frac{k_{\bot}^2}{|q_u eB|}\right)L_{k-1}\left(2\frac{k_{\bot}^2}{|q_d eB|}\right)+L_{p-1}\left(2\frac{k_{\bot}^2}{|q_u eB|}\right)L_k\left(2\frac{k_{\bot}^2}{|q_d eB|}\right)\bigg]\nonumber\\
&&+64k_{\bot}^2L_{p-1}^1\left(2\frac{k_{\bot}^2}{|q_u eB|}\right)L_{k-1}^1\left(2\frac{k_{\bot}^2}{|q_d eB|}\right)\Bigg{\}}.
\eea
Here, the functions $I_1'$, $I_1''$ and $I_2'$ are represented in appendix \ref{Integralk0}. Because of electric charge difference, we could not get a simple form as in Eq.(\ref{deltalandau}) for $\sigma$ meson, so we perform the numerical calculation for the integral of $k_{\bot}$. Also we will do the same operation in the diquark case.

 Similarly, we use the following gap equation to obtain the mass of $\pi$ meson
 \bea\label{Gappi}
 1-2G_S\Pi_{\pi^+/\pi^0}(q_{\bot},q_{\parallel})=0.
 \eea

\subsection{The diquark in the scalar channel}
  For the diquark channel, the interaction term in the Lagrangian $\cal{L}_{\text{I}}$ consists the term of the form $ (\bar{\psi}A\bar{\psi}^T)(\psi^{T}B\psi)$ and the A,B are matrices antisymmetric in Dirac, isospin and color indices. Here we only consider the scalar channel in the  color $\bar{3}$ channels, the interaction term $\cal{L}_{\text{I,D}}$ is given by \cite{Ishii:1995bu,Huang:2002zd,Huang:2004ik}
  \bea
  {\cal{L}}_{\text{I,D}}=G_D[(i {\bar{\psi}}^C  \varepsilon  \epsilon^{b} \gamma_5 \psi )
   (i {\bar{\psi}} \varepsilon \epsilon^{b} \gamma_5 \psi^C)],
  \eea
with $\psi^C=C {\bar{\psi}}^T$, ${\bar{\psi}}^C=\psi^T C$ are charge-conjugate spinors, $C=i \gamma^2 \gamma^0$ is the charge
conjugation matrix (the superscript $T$ denotes the transposition operation), the quark field $\psi \equiv \psi_{i\alpha}$ with $i=1,2$ and $\alpha=1,2,3$ is a flavor
doublet and color triplet, as well as a four-component Dirac spinor, $(\varepsilon)^{ik} \equiv \varepsilon^{ik}$, $(\epsilon^b)^{\alpha \beta} \equiv \epsilon^{\alpha \beta b}$ are totally antisymmetric tensors in the flavor and color spaces. In this work, we choose $G_D=\frac{3}{4}G_S$.

Then we introduce the auxiliary diquark fields $\Delta^b$ and $\Delta^{*b}$
\begin{eqnarray}
\Delta^b \sim i {\bar{\psi}}^C \varepsilon \epsilon^{b}\gamma_5 \psi, \ \
\Delta^{*b} \sim i {\bar{\psi}}  \varepsilon  \epsilon^{b} \gamma_5 \psi^C,\ \
\end{eqnarray}
which are color antitriplet and (isoscalar) singlet under the chiral $SU(2)_L \times SU(2)_R$ group.

The gap equation for the diquark in the color antitriplet scalar channel is given by~\cite{Ishii:1995bu}
  \bea
  1+2G_{D}\Pi_{\Delta^b}(q_{\bot},q_{\parallel})=0,
  \eea
where
\bea
&&\Pi_{\Delta^b}(q_{\bot},q_{//})\delta_{A'A}=-i\int\frac{d^4k}{(2\pi)^4}\times3tr_{\Delta^{b}}[\gamma_5\widetilde{S}(-d,k)\gamma_5\widetilde{S}(u,k+q)]+(u\Leftrightarrow d)\nonumber\\
&=&i\int\frac{d^4k}{(2\pi)^4}\times3\exp\left(-\frac{k^2_{\bot}}{|q_deB|}\right)\exp\left(-\frac{k^2_{\bot}}{|q_ueB|}\right)\nonumber\\
&&\sum_{p,k=0}^{\infty}(-1)^{p+k}\frac{1}{(k_0^2-k_3^2-M^2-2|q_deB|k)(p_0^2-k_3^2-M^2-2|q_ueB|p)}\nonumber\\
&&\Bigg{\{}8(-p_{||}\cdot k_{||}+M^2)\bigg(L_k\left(2\frac{k^2_{\bot}}{|q_deB|}\right)L_p\left(2\frac{k^2_{\bot}}{|q_ueB|}\right)+
 L_{k-1}\left(2\frac{k^2_{\bot}}{|q_deB|}\right)L_{p-1}\left(2\frac{k^2_{\bot}}{|q_ueB|}\right)\bigg)\nonumber\\
 &&+64k_{\bot}^2L_{k-1}^1\left(2\frac{k^2_{\bot}}{|q_deB|}\right)L_{p-1}^1\left(2\frac{k^2_{\bot}}{|q_ueB|}\right)\Bigg{\}}+(u\Leftrightarrow d).
\eea
By introducing the functions $I_1',I_2'$ defined in Appendix, we can express the diquark loop polarization function as:
 \bea
\Pi_{\Delta^b}(q_{\bot},q_{//})\delta_{A'A}=
 &=&24i\int\frac{d^3\mathbf{k}}{(2\pi)^3}\exp\left(-\frac{k^2_{\bot}}{|q_deB|}\right)\exp\left(-\frac{k^2_{\bot}}{|q_ueB|}\right)\nonumber\\
&&\sum_{p,k=0}^{\infty}(-1)^{p+k}\Bigg{\{}\left[-\frac{1}{2}(I_1'+I_1'')-(|q_d eB|k+|q_u eB|p-\frac{1}{2}M^2_{\Delta^{b}})I_2'\right]\nonumber\\&&\bigg(L_k\left(2\frac{k^2_{\bot}}{|q_deB|}\right)L_p\left(2\frac{k^2_{\bot}}{|q_ueB|}\right)+
 L_{k-1}\left(2\frac{k^2_{\bot}}{|q_deB|}\right)L_{p-1}\left(2\frac{k^2_{\bot}}{|q_ueB|}\right)\bigg)\nonumber\\
 &&+64k_{\bot}^2L_{k-1}^1\left(2\frac{k^2_{\bot}}{|q_deB|}\right)L_{p-1}^1\left(2\frac{k^2_{\bot}}{|q_ueB|}\right)\Bigg{\}}+(u\Leftrightarrow d).
\eea
 Here, $p=k+q_{\Delta^b}$ and we assume $q_{\Delta^b}^{\mu}=(M_{\Delta^b},\mathbf{0})$ (i.e., in the rest frame of diquark).

\section{NUMERICAL RESULTS}
\label{Numerical}
 In the numerical calculation, we use the soft cut-off functions\cite{Frasca:2011zn}
 \bea
&&f_\Lambda=\sqrt{\frac{\Lambda^{10}}{\Lambda^{10}+\mathbf{k}^{2*5}}}, \\
&&f_{\Lambda,eB}^k=\sqrt{\frac{\Lambda^{10}}{\Lambda^{10}+(k_3^2+2|Q eB|k)^5}},
\eea
for zero magnetic field and nonzero magnetic field respectively. In order to reproduce the pion mass $M_{\pi}=140$ MeV, pion decay constant $f_{\pi}=92.3$ MeV, constituent quark mass $M=336$ MeV in the vacuum, we choose the following model parameters: $\Lambda=616$ MeV, $G_S\Lambda^2=2.02$  and the current quark mass $m_0$ is $5$ MeV.

In  Fig.\ref{fig:quarkmass}, we show the magnetic field dependence of quark mass with different fixed temperatures and chemical potentials. We can see that the quark mass increases when the magnetic field strength increases, which is the magnetic catalysis effect obviously. It is noticed that in the regular NJL model, there is no mechanism for the inverse magnetic catalysis around the critical temperature region, therefore, all results in this work are taken below the critical temperature.

 \begin{figure}[!thb]
\centerline{\includegraphics[width=8cm]{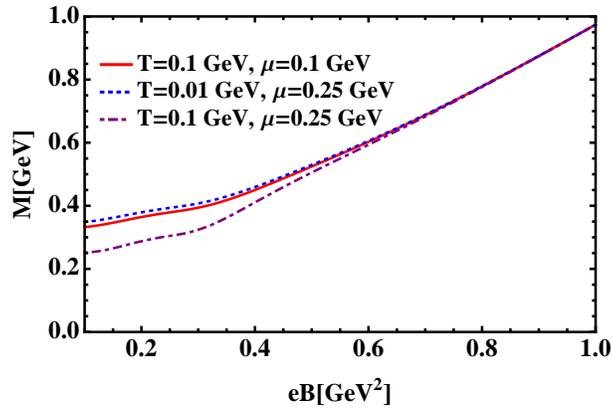}}
\caption{The $eB$ dependence of the constituent quark  mass $M$ with fixed different temperature and chemical potential. }
\label{fig:quarkmass}
\end{figure}

\subsection{Numerical results for neutral $\sigma$ and $\pi^0$}

 We firstly check the mass behavior of charge neutral $\sigma$ and $\pi^0$ under magnetic fields.

The numerical result for the mass square of the $\sigma$ meson is shown in Fig.\ref{fig:sigmamass}, and the constant mass of the point-particle model is also listed for comparison. It is observed that when the quark-loop polarization under magnetic fields is considered, similar to the behavior of the quark mass, the mass of the $\sigma$ meson increases with magnetic field for different temperatures and chemical potentials. Comparing with the constant mass of $\sigma$ meson as a point-particle, we find that the quark-loop contribution or quark polarization effect is very essential.  On the other hand, at zero temperature and below the critical chemical potential, we find that the chemical potential does not affect the mass of the $\sigma$ meson, and for $\mu=0$ MeV case, the mass of $\sigma$ meson decreases when temperature increases at fixed magnetic field.

 \begin{figure}[h!]
 \centering
 \includegraphics[width=7cm]{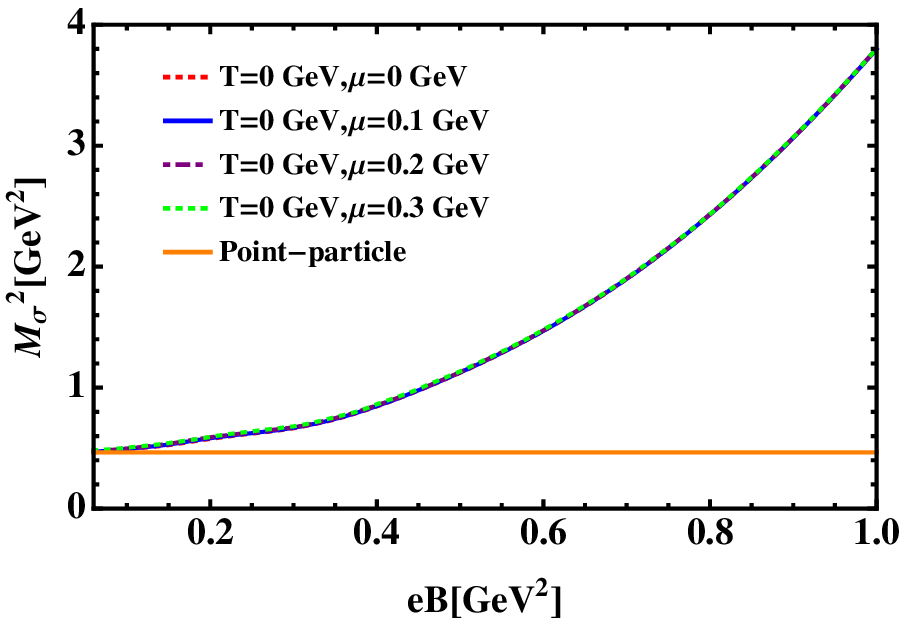}
 \includegraphics[width=7cm]{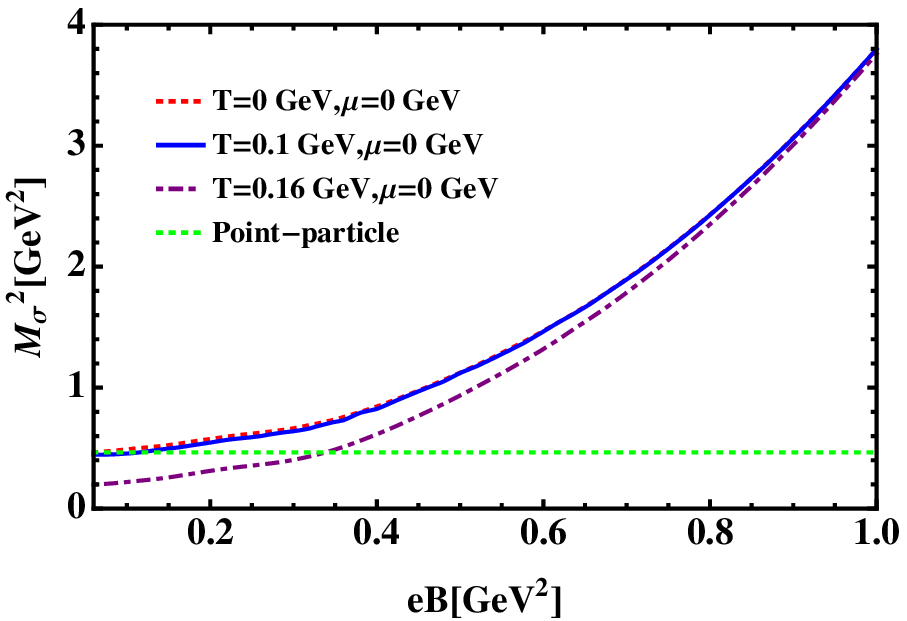}
\caption{The magnetic dependence of the mass square of $\sigma$ meson. (Left) The mass square of $\sigma$ meson with fixed $T=0$ MeV with different $\mu$. (Right) The mass square of $\sigma$ meson with fixed $\mu=0$ MeV with different $T$.}
\label{fig:sigmamass}
\end{figure}

In Fig.\ref{fig:neutrapimass}, we show the magnetic field dependence of the mass square for neutral pion $\pi^0$, and compare with the result from the point-particle model.  The neutral pion mass keeps as a constant as a function of the magnetic field in the point-particle model. It is observed that in the vacuum when $T=0,\mu=0$, the quark-loop contribution has tiny effect on neutral pion mass, which almost keeps a constant with the increasing of the magnetic field. This is preserved by the remnant symmetry and $\pi^0$ is the only pseudo Goldstone boson in the system. But even it is a tiny effect, we still can see that the $\pi^0$ mass decreases a little firstly and then increases a little with the magnetic field, which is in agreement with the lattice result in Ref.\cite{Bali:2015vua}. At zero temperature when $T=0$ MeV, the mass of $\pi^0$ meson does not change with different chemical potentials in the chiral symmetry breaking phase, and at zero baryon density,  the increasing temperature lowers the mass of neutral $\pi^0$ from vacuum mass $140 {\rm MeV}$ to almost 0 around the critical temperature.

\begin{figure} [!h]
\centering
\includegraphics[width=7cm]{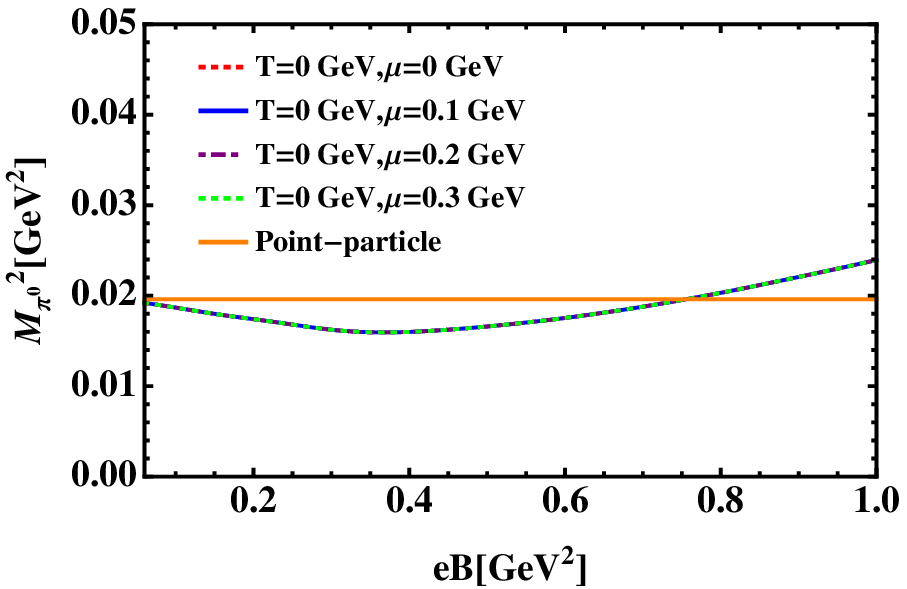}
\includegraphics[width=7cm]{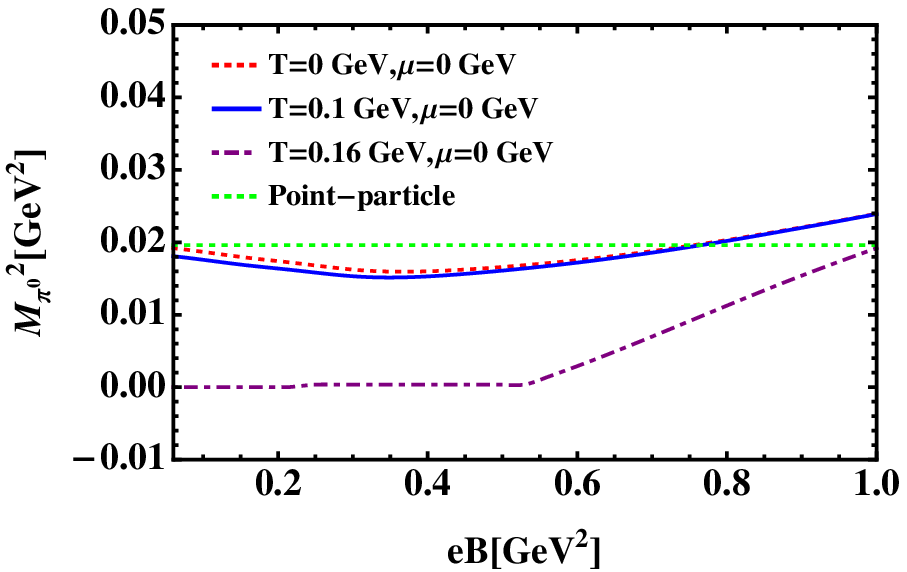}
\caption{The magnetic field dependence of the mass square of neutral $\pi^0$ meson. (Left) The mass square of neutral $\pi^0$ meson with fixed $T=0$ MeV with different $\mu$. (Right) The mass square of neutral $\pi^0$ meson with fixed $\mu=0$ MeV with different  $T$.}
\label{fig:neutrapimass}
\end{figure}

In the point-particle model, the mass of a neutral meson either for scalar $\sigma$ or peudo-scalar $\pi^0$ should keep a constant as the function of the magnetic field. However, when the polarized quark-loop effect is considered, the results for scalar $\sigma$ or peudo-scalar $\pi^0$ are quite different. Similar to the quark mass, the mass of neutral scalar $\sigma$ linearly rises as the magnetic field and one has $M_{\sigma}(eB)\simeq 2 M(eB)$. However, as the only pseudo Goldstone meson, the neutral $\pi^0$ mass keeps as an almost constant value, which is preserved by the remnant symmetry of the system.

\subsection{Numerical results for charged $\pi^{\pm}$ and $\Delta^b$}

Then we analyze the mass behavior for charged $\pi^{\pm}$ with electric charge $\pm1$ and $\Delta^b$ with electric charge $1/3$ under the magnetic field.

 By solving the gap equation Eq.(\ref{Gappi}) numerically,  the mass square of charged $\pi^+$ meson is given in Fig.\ref{fig:cahrgepimass} where we also list the result of the point-particle model. The mass of charged $\pi^+$ increases with increasing magnetic field in different temperature $T$ and chemical potential $\mu$. The chemical potential $\mu$ has no effect on the mass of the charged $\pi^+$ with zero temperature in the chiral symmetry breaking phase, the increasing temperature shows tiny effect on the mass of the $\pi^+$ meson. Obviously, the quark-loop contribution plays an important role by comparing with the results of the one-loop polarization function and the point-particle model.

 \begin{figure}[h!]
 \centering
\includegraphics[width=7cm]{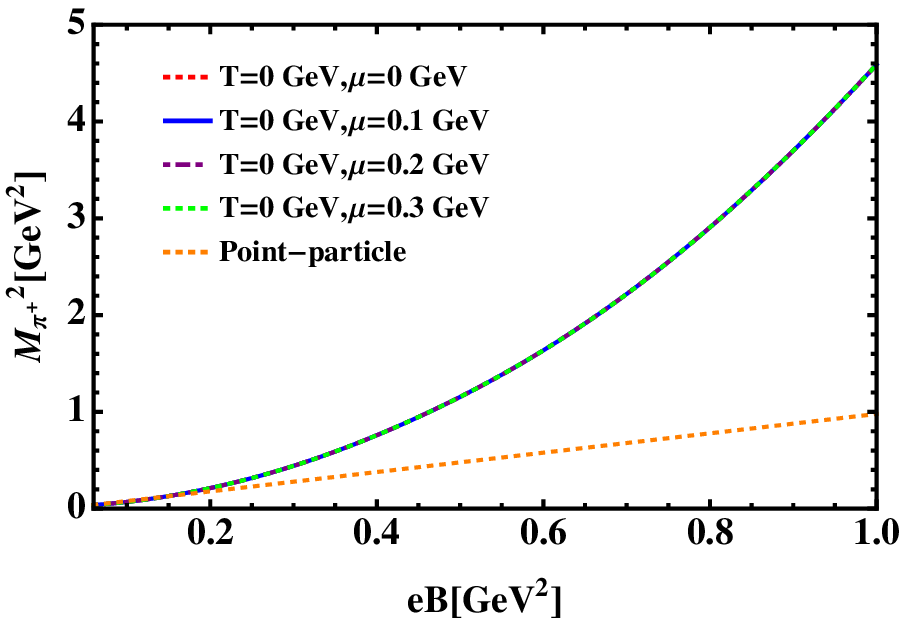}
\includegraphics[width=7cm]{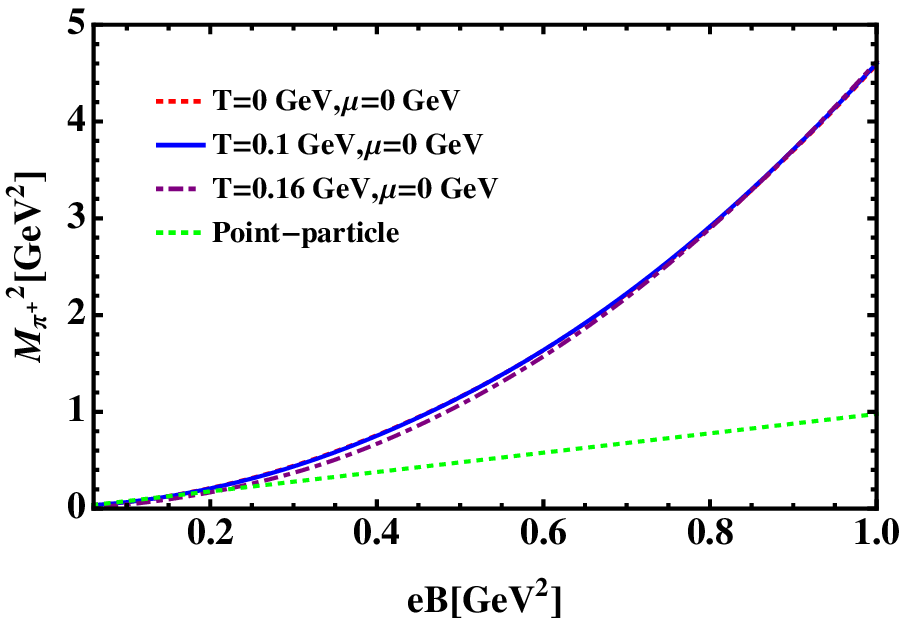}
\caption{The mass square of charged $\pi^+$ meson in the magnetic field. (Left) The mass square of charged $\pi^+$ meson with fixed  $T=0$ MeV with different $\mu$. (Right) The mass  of charged $\pi^+$ meson with fixed $\mu=0$ MeV with different  $T$.}
\label{fig:cahrgepimass}
\end{figure}

 \begin{figure}[h!]
 \centering
\includegraphics[width=7cm]{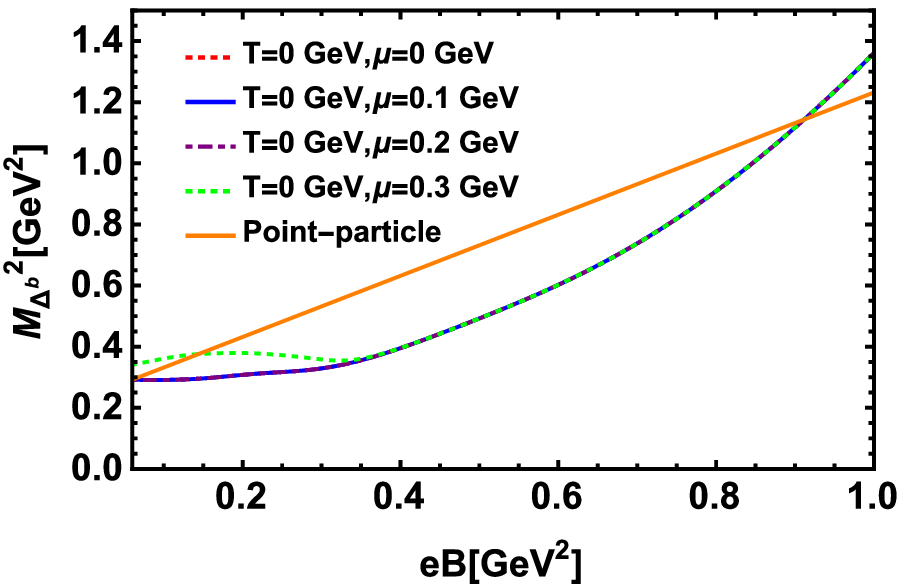}
\includegraphics[width=7cm]{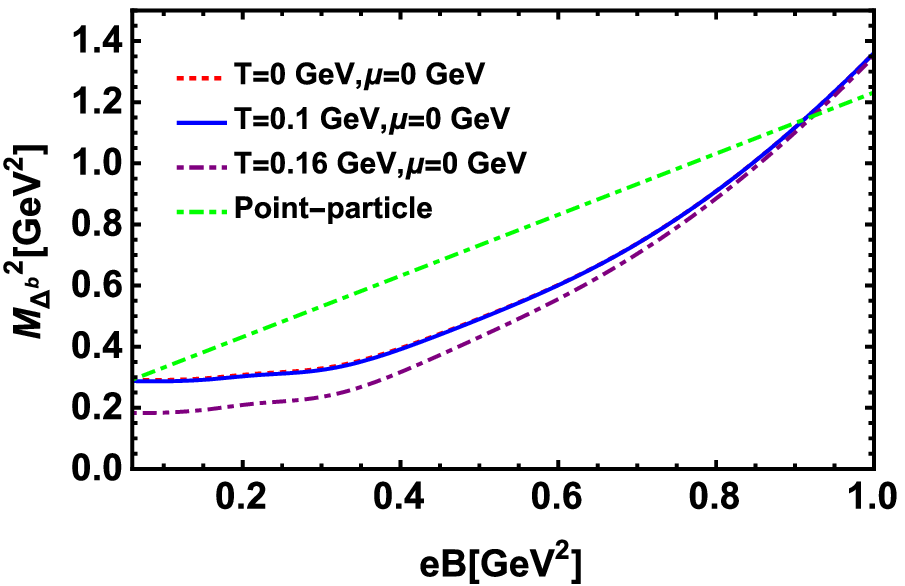}
\caption{The mass square of diquark in the scalar channel in the magnetic field. (Left) The mass  of diquark in the scalar channel with fixed $T=0$ MeV
for different  $\mu$. (Right) The mass  of diquark in the scalar channel with fixed $\mu=0$ MeV for different  $T$. }
\label{fig:diquarkmass}
\end{figure}

 We also investigate the mass square of scalar diquark in color anti-triplet channel in magnetic field with different temperature and chemical potential in Fig.\ref{fig:diquarkmass}. The mass of scalar diquark increases with the magnetic field. At zero temperature, the chemical potential affects the mass of diquark in the scalar channel slightly. Moreover, at zero chemical potential, the mass of diquark in the scalar channel decreases with the increasing temperature. Generally, the one-loop contribution has important effect on the mass of scalar diquark. However, it is important to notice that different from charged $\pi^{\pm}$, including the quark-loop contribution, the mass of the scalar diquark increases more slowly with magnetic fields comparing with the point-particle model result when $eB<0.9~ \text{GeV}^2$, and only at high magnetic field when $eB>0.9~ \text{GeV}^2$, the mass of scalar diquark increases faster with magnetic fields comparing with the result of the point-particle model.

\subsection{Quark-loop contribution}

It is observed that except for the charge neutral $\pi^0$, which is the only pseudo Goldstone boson of the system preserved by the remnant symmetry of QCD$\times$QED, for all other scalar mesons including the charge neutral $\sigma$, $\pm1$ charged pseudo scalar $\pi^{\pm}$ and $1/3$ charged scalar diquark $\Delta^b$, their masses including the one quark-antiquark loop contribution under magnetic fields are quite different from the point-particle results. We explicitly show the one quark-loop contribution to $\sigma$, $\pi^{\pm}$ and $\Delta^b$ as the function of the magnetic field at $T=0,\mu=0$ in Fig.\ref{fig:oneloopcontribution}. The one quark-loop contribution to the charge neutral $\sigma$ and charged $\pi^{\pm}$ increases essentially with the the increase of magnetic fields due to the magnetic catalysis of the polarized quarks. At strong magnetic fields, when $eB> 0.5 {\rm GeV}^2$, the one quark-loop contribution to the mass of charge neutral $\sigma$ meson becomes more essential than the charged  $\pi^{\pm}$, and at $eB=1 {\rm GeV}^2$, the one quark-loop contribution to the mass of charge neutral $\sigma$ meson can reach 8 times of the point-particle results. However, the one quark-quark loop contribution to the scalar diquark mass is negative and  less than 50$\%$ of the point-particle result below $eB<0.9 {\rm GeV}^2$.

\begin{figure}[h!]
\centerline{\includegraphics[width=8cm]{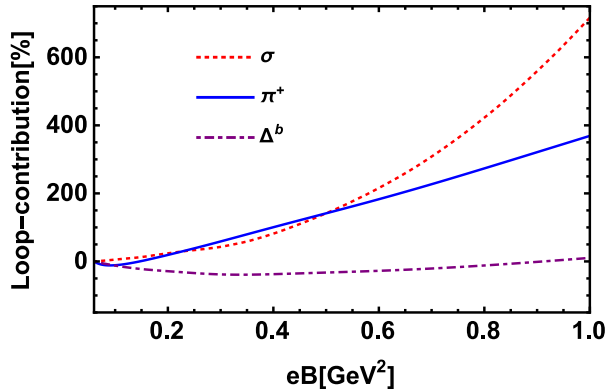}}
\caption{The one quark-loop contribution to the mass of the $\sigma$, $\pi^+$ and the diquark $\Delta^b$ in the scalar channel. }
\label{fig:oneloopcontribution}
\end{figure}

\section{Discussion and Conclusion}
\label{Conclusion}
In this paper, we have studied the masses of charge neutral scalar $\sigma$ and pseudo-scalar $\pi^0$, charged pseudo-scalar meson $\pi^{\pm}$ and scalar diquark $\Delta^b$ in an external magnetic field in the NJL model, and the mesons and diquark are constructed by summing up infinite quark-loop chains in the random phase approximation (RPA), the quark-loop of mesons polarization function is calculated to the leading order of $1/N_c$.

It is found that the mass of the charge neutral $\pi^0$ keeps almost constant under the magnetic field and the one quark-loop contribution to $\pi^0$ is tiny. This is because that $\pi^0$ is the only pseudo Goldstone boson of the system preserved by the remnant symmetry of QCD$\times$QED. From the point-particle model, the mass of the charge neutral scalar $\sigma$ should keep as a constant value under the magnetic field, however, considering the one quark-loop contribution, the $\sigma$ mass is around two times quark mass and increases with the magnetic field due to the magnetic catalysis effect. This is an typical example showing that the polarized quark structure cannot be neglected when we consider the meson properties under magnetic field, and the quark-loop contribution to $\sigma$ mass plays dominant role at strong magnetic field, e.g., the quark-loop contribution to $\sigma$ mass can reach 8 times of the point-particle results at $eB=1 {\rm GeV}^2$.

For the charged particles,  the contribution from the one quark-loop are quite different to the masses of pseudo-scalar $\pi^{\pm}$ and scalar diquark $\Delta^b$ under magnetic fields. The one quark-antiquark loop contribution to the charged $\pi^{\pm}$ increases essentially with the the increase of magnetic fields due to the magnetic catalysis of the polarized quarks.  However, the one quark-quark loop contribution to the scalar diquark mass is negative and  less than 50$\%$ of the point-particle result below $eB<0.9 {\rm GeV}^2$.

As mentioned in the Introduction that one of the motivations of this work is to check whether the method we used to calculate the charged vector meson \cite{Liu:2014uwa,Liu:2015pna,Liu:2016vuw} is correct. From the results of pseudo-scalar $\pi$ by using the same method, we are confident that the results of charged vector meson in \cite{Liu:2014uwa,Liu:2015pna,Liu:2016vuw} is correct at least in the framework of the NJL model. Firstly, we have reproduced the result for the charge neutral $\pi^0$, its mass is almost a constant under the magnetic fields, and this is preserved by the remnant symmetry and $\pi^0$ is the only pseudo Goldstone boson in the system. Secondly, the charged pion mass increases with the magnetic field, which is in agreement with lattice results in \cite{Bali:2015vua,Luschevskaya:2015cko}.

At last, we have to mention that when this paper is almost finished, there is a similar paper published~\cite{Wang:2017vtn}, where the authors have pointed out that the method we are using in \cite{Liu:2014uwa,Liu:2015pna,Liu:2016vuw} only take the translation invariance part of the Schwinger phase in quark propagator. In fact, we only keep the first order of local expansion for the charged field and the phase in our paper, which leads to a cancellation of the translation variance part. Compare the result in both papers, we can see, the first-order expansion already gives a very good result and physical quantities.

\begin{acknowledgments}
We thank useful discussion with S.J.Mao. The work of M.Huang is supported by the NSFC under Grant Nos. 11725523, 11735007 and 11261130311(CRC 110 by DFG and NSFC). L.Yu is supported by the NSFC under Grant No. 11605072 and the Seeds Funding of Jilin University.
\end{acknowledgments}

\appendix

\section{Integrals}
\label{Integralk0}
We have introduced the integrals of $k_0$ for different channels, for the neutral meson,
 \bea
  I_1&=&\int\frac{dk_0}{2\pi}\frac{1}{k_0^2-\omega^2},\nonumber\\
 I_2&=&\int\frac{dk_0}{2\pi}\frac{1}{(k_0^2-\omega^2)((k_0+M_{\pi^0/\sigma})^2-\omega^2)},
 \eea
 with $\omega=\sqrt{2|q_feB|k+k_3^2+M^2}$.  For the charged $\pi^+$ meson and the diquark in the scalar channel,
 \bea
 &&I'_1=\int\frac{dk_0}{2\pi}\frac{1}{k_0^2-\omega_{u,p}^2}, \\
&&I_1^{''}=\int\frac{dk_0}{2\pi}\frac{1}{k_0^2-\omega_{d,k}^2}, \\
&&I_2'=\int\frac{dk_0}{2\pi}\frac{1}{((k_0+q_0)^2-\omega_{u,p}^2)(k_0^2-\omega_{d,k}^2)}.
\eea
with $\omega_{u,p}=\sqrt{k_3^2+2|q_u eB|p+M^2}$, $\omega_{d,k}=\sqrt{k_3^2+2|q_d eB|k+M^2}$. Following the Ref.\cite{Rehberg:1995nr}, the integral of $k_0$ can be replaced by the matsubara sum,$\int\frac{dk_0}{2\pi}(....)=iT\sum_{m=-\infty}^{\infty}(....)$, and we can obtain
\bea
iI_1&=&-\left[\frac{n_f(\omega-\mu)+n_f(\omega+\mu)-1}{2\omega}\right],\\
iI_2&=&-\left[\frac{n_f(\omega-\mu)}{2\omega}\frac{1}{(\omega+M_{\pi_0/\sigma})^2-\omega^2}-\frac{n_f(-\omega-\mu)}{2\omega}\frac{1}{(-\omega+M_{\pi^0/\sigma})^2-\omega^2}\right.\nonumber\\
&&\left.+\frac{n_f(\omega-
\mu)}{2\omega}\frac{1}{(-M_{\pi^0/\sigma}+\omega)^2-\omega^2}-\frac{n_f(-\omega-\mu)}{2\omega}\frac{1}{(-M_{\pi^0/\sigma}-\omega)^2-\omega^2}\right],\\
iI_1'&=&-\left[\frac{n_f(\omega_{u,p}-\mu)+n_f(\omega_{u,p}+\mu)-1}{2\omega_{u,p}}\right],\\
iI_1^{''}&=&-(\frac{n_f(\omega_{d,k}-\mu)+n_f(\omega_{d,k}+\mu)-1}{2\omega_{d,k}}),\\
iI_2'&=&-\left[\frac{n_f(\omega_{d,k}-\mu)}{2\omega_{d,k}}\frac{1}{(\omega_{d,k}+M_{\pi^+/\Delta^b})^2
-\omega^2_{u,p}}-\frac{n_f(-\omega_{d,k}-\mu)}{2\omega_{d,k}}\frac{1}{(-\omega_{d,k}+M_{\pi^+/\Delta^b})^2-\omega^2_{u,p}}\right.\nonumber\\
&&\left.+\frac{n_f(\omega_{u,p}-
\mu)}{2\omega_{u,p}}\frac{1}{(-M_{\pi^+/\Delta^b}+\omega_{u,p})^2-\omega^2_{d,k}}
-\frac{n_f(-\omega_{u,p}-\mu)}{2\omega_{u,p}}\frac{1}{(-M_{\pi^+/\Delta^b}-\omega_{u,p})^2-\omega^2_{d,k}}\right],
\eea
with $n_f(x)=\frac{1}{1+e^{\frac{x}{T}}}$.

 \end{document}